\documentclass[aps,preprintnumbers,amsmath,twocolumn,groupedaddress,superscriptaddress,notitlepage,nofootinbib,prd]{revtex4-1}
\usepackage[utf8]{inputenc}
\usepackage{indentfirst}
\usepackage{amssymb}
\usepackage{amsmath,latexsym}
\usepackage{graphicx}
\usepackage{float}
\usepackage[margin=1.0in]{geometry}
\usepackage[mathscr]{euscript}
\usepackage{color}
\usepackage[dvipsnames]{xcolor}
\usepackage{soul}
\usepackage{slashed}
\usepackage{feynmf}
\usepackage{braket}
\usepackage{comment}
\usepackage{xcolor}

\usepackage{dsfont}
\newcommand{\bbid}{\mathds{1}}
\newcommand{\beq}{\begin{equation}}
\newcommand{\eeq}{\end{equation}}
\newcommand{\BE}{\begin{equation}}
\newcommand{\EE}{\end{equation}}
\newcommand{\bea}{\begin{eqnarray}}
\newcommand{\eea}{\end{eqnarray}}

\long\def\beqs#1\eeqs{\beq\begin{split} #1 \end{split}\eeq}

\newcommand{\me}{\mathrm{e}}
\newcommand{\eps}{\epsilon}
\newcommand{\veps}{\varepsilon}

\newcommand{\mcal}{\mathcal}

\newcommand{\smallfrac}[2]{{\textstyle\frac{#1}{#2}}}
\newcommand{\avg}[1]{\langle #1 \rangle}
\newcommand{\mrm}{\mathrm}
\newcommand{\mbb}{\mathbb}

\newcommand{\dd}{\mathrm{d}}

\newcommand{\what}{\widehat}

\definecolor{MyRed}{RGB}{153,0,13}

\usepackage[colorlinks=true,backref=true,linktocpage=true,citecolor=MyRed,urlcolor=MyRed,linkcolor=MyRed,pdfpagemode=UseOutlines]{hyperref}

\hypersetup{%
  bookmarksnumbered=true,
  pdftitle = {},
  pdfsubject = {},
  pdfauthor = {},
  pdfkeywords = {}
}

\usepackage{microtype}

\usepackage{pgfplotstable,array,siunitx}
\usepackage{multirow}
\usepackage{booktabs}

\usepackage{tikz}

\preprint{INT-PUB-20-030}

\def\av#1{ \left\langle #1 \right\rangle }

\newcommand{\nn}{\nonumber}
\newcommand{\eq}[1]{Eq.~(\ref{#1})}
\newcommand{\fig}[1]{Fig.~\ref{#1}}

\begin{document}
\title{On the Infinite Variance Problem in Fermion Models}

\author{Andrei Alexandru}
\email{aalexan@gwu.edu}
\affiliation{Department of Physics,
The George Washington University, Washington, DC  20052}
\affiliation{Department of Physics,
University of Maryland, College Park, MD 20742}

\author{Paulo F. Bedaque}
\email{bedaque@umd.edu}
\affiliation{Department of Physics,
University of Maryland, College Park, MD 20742}

\author{Andrea Carosso}
\email{acarosso@gwu.edu}
\affiliation{Department of Physics,
The George Washington University, Washington, DC  20052}

\author{Hyunwoo Oh}
\email{hyunwooh@umd.edu}
\affiliation{Department of Physics,
University of Maryland, College Park, MD 20742}

\preprint{}

\date{\today}

\begin{abstract}

Monte Carlo calculations of fermionic systems with continuous auxiliary fields frequently suffer from a diverging variance for fermionic observables. If the simulations have an infinite variance problem, one cannot estimate observables reliably even with an arbitrarily large number of samples. In this paper, we explore 
a method to solve this problem using sampling based on the distribution of a system with an extra time-slice. The necessary reweighting
factor is computed both perturbatively and through a secondary Monte Carlo.
We show that the Monte Carlo reweighting coupled to the use of an unbiased estimator of the reweighting factor leads to a method that eliminates the infinite variance problem at a very small extra cost.
 We compute the double occupancy in the Hubbard model at half-filling to demonstrate  the method and compare the results to well established results obtained by other methods. 

\end{abstract}

\maketitle

\section{Introduction}

Lattice fermion models are ubiquitous in many areas of physics, from condensed matter to lattice QCD. Frequently, no analytic methods are available to study them, so stochastic methods are used instead. But fermions pose a number of special challenges to Monte Carlo calculations, among them the infinite variance problem we address in this paper.
The Hubbard model is an example of a lattice fermion model and has attracted much attention since it not only exhibits a rich phase structure as the temperature and doping are varied (including  antiferromagnetic, superconducting, and non-Fermi liquid regions) but also because it is supposed to be intimately connected to the physics of cuprate high-T$_c$ superconductors.

A common strategy for simulating the Hubbard model is to introduce an auxiliary bosonic variable via the Hubbard-Stratonovich transformation, which allows for a Monte Carlo simulation over those variables (instead of the original fermionic ones) to be  carried out.
Although \textit{discrete} auxiliary variables are most commonly used, there are at least two reasons for using \textit{continuous} variables instead. First, away from half-filling, the action of the auxiliary field becomes complex and the model develops a sign problem, making a Monte Carlo simulation nearly impossible. The methods used to bypass the sign problem, such as the complex Langevin method, are all based on an analytic continuation of the auxiliary variables \cite{Fukuma:2019wbv,Mori:2017pne,Wynen:2020uzx,Rodekamp:2022xpf,Endres:2011jm} (see \cite{Alexandru:2020wrj} for a review), all require the use of continuous variables. 
Secondly, the very successful hybrid Monte Carlo algorithm, which speeds up Monte Carlo calculations in other models, also requires continuous auxiliary fields \cite{Wynen:2018ryx,Krieg:2018pqh,Ulybyshev:2019fte}.
However, the desirability of continuous auxiliary variables is tempered by the fact that their use typically leads to
large (or even infinite) variance for interesting observables, again making Monte Carlo calculations extremely difficult. A manifestation of this problem also occurs in lattice QCD for discretizations that encounter ``exceptional'' gauge configurations~\cite{Luscher:1996ug}.
Note that while for QCD this problem was detected in the context of quenched simulations and attributed to neglecting the determinant in the sampling weight, similar divergences crop up in dynamical simulations~\cite{DelDebbio:2005qa} which lead to instabilities in the HMC algorithm. The solution for these instabilities~\cite{Luscher:2008tw,Miao:2011sx,Bruno:2014jqa} is similar to the solution we explore here: use sampling with a positive weight that has no zeros in the field space together with a reweighting.

A number of methods have been proposed to deal with the infinite variance problem \cite{Shi:2016,Ulybyshev:2021fsb,Yunus:2022pto,Nicholson:2012xt}. The origin of the problem is that some auxiliary field configurations with small statistical weights contribute significantly to observables. A way of dealing with this issue, proposed in
\cite{Shi:2016}, 
is to sample the configurations according to a \textit{modified} distribution so that small statistical weights and large contributions to the observables do not occur for the same configurations.
The proposal was made in the context of auxiliary-field Quantum Monte Carlo with noncompact auxiliary fields at zero temperature, and consists in a reweighting with respect to a probability density involving \textit{one more time-slice} than the system being considered.

In this work, we adapt the extra time-slice method to be suitable for standard Monte Carlo simulations of Euclidean time path integrals over \textit{compact} auxiliary fields, and show how it is implemented for two discretizations of the action of the system.
The implementation proposed in \cite{Shi:2016}, however, requires that one compute fermion contractions of $\exp(-\eps H)$ to obtain the reweighting factor, where $H$ is the Hamiltonian operator of the system under investigation. Two options present themselves: compute several terms in an expansion in $\eps$, or estimate it nonperturbatively using Monte Carlo by writing it as an integral with respect to a known probability distribution. Since the first option introduces new $O(\eps^k)$ errors, and quickly becomes difficult to calculate at orders $k> 2$, the second option is preferable. The latter option amounts to a stochastic estimate of $1/F$, where $F$ is a quantity estimated by Monte Carlo---but the estimator $1/\overline F$ for this quantity is known to be biased~\cite{Moka:2019}. We therefore implement an unbiased estimator of the reweighting factor and show the improvement obtained by this method.

In what follows, we first define our conventions for the testbed model used in our investigation, namely, the Hubbard model at half-filling. We then review the origins of the infinite variance problem and display its symptoms in a Monte Carlo simulation. The extra time-slice method is then defined, together with the two approaches to computing the requisite reweighting factor. Lastly, we describe the unbiased estimation of the reweighting factor.

\section{Model}

We denote an electron creation operator at site $x$ by $\hat \psi_{a,x}^\dag,$ where $a=\uparrow,\downarrow$ is the spin. The grand canonical Hamiltonian of the Hubbard model on a square lattice is then given by
\beq
    \begin{aligned}
    H &=- \kappa \sum_{\av{x,y}}(\hat{\psi}^{\dagger}_{\uparrow,x}\hat{\psi}_{\uparrow,y}+\hat{\psi}^{\dagger}_{\downarrow,x}\hat{\psi}_{\downarrow,y}) \\
    & + U \sum_{x}(\hat{\psi}^{\dagger}_{\uparrow,x}\hat{\psi}_{\uparrow,x}-\smallfrac{1}{2})
     (\hat{\psi}^{\dagger}_{\downarrow,x}\hat{\psi}_{\downarrow,x}-\smallfrac{1}{2}) \\
     & - \mu \sum_{x}(\hat{\psi}^{\dagger}_{\uparrow,x}\hat{\psi}_{\uparrow,x}+\hat{\psi}^{\dagger}_{\downarrow,x} \hat{\psi}_{\downarrow,x}-1).
    \end{aligned}
\eeq
Here, we have used parameters which yield half-filling at $\mu=0$. The summation in the first term is over sites $\av{x,y}$ that are nearest neighbors. On a lattice with an even number of sites, we define $\hat{\psi}_{1,x}\equiv \hat{\psi}_{\uparrow,x},$ $\hat{\psi}_{2,x}\equiv(-1)^x\hat{\psi}^{\dagger}_{\downarrow,x}$, so that, in terms of the new variables, the Hamiltonian becomes
\beq
    \begin{aligned}
        H = & - \kappa \sum_{\av{x,y}}(\hat{\psi}^{\dagger}_{1,x}\hat{\psi}_{1,y}+\hat{\psi}^{\dagger}_{2,x}\hat{\psi}_{2,y}) \\
        & + \frac{U}{2}\sum_{x}(\hat{\psi}^{\dagger}_{1,x}\hat{\psi}_{1,x}-\hat{\psi}^{\dagger}_{2,x}\hat{\psi}_{2,x})^2 \\
        & - \mu\sum_{x}(\hat{\psi}^{\dagger}_{1,x}\hat{\psi}_{1,x}-\hat{\psi}^{\dagger}_{2,x}\hat{\psi}_{2,x})
    \end{aligned}
\eeq
up to a constant shift.

Using a Hubbard-Stratonovich transformation, the expectation value of an observable $\mcal{O}$ in a thermal ensemble at temperature $T$ has a representation as an integral over an auxiliary bosonic field defined on the same square lattice:
\beq\label{eq:expectation}
\langle \mathcal{O}\rangle=
\frac{
\int [\dd\phi] \mathcal{O}(\phi) \; \me^{-S_0(\phi)} \det\mbb M(\phi)
}
{\int [\dd\phi] \; \me^{-S_0(\phi)} \det\mbb M(\phi)
},
\eeq
where the auxiliary field action $S_0(\phi)$ is
\beq \label{free-action}
S_0(\phi) = -\beta \sum_{x,t}  \cos\phi_{t,x}.
\eeq
As described in the appendix, one can derive this using a Trotterization of the partition function $Z={\rm Tr}\,\exp(-H/T)\approx{\rm Tr}\,[\exp(-\epsilon H)]^N$. 
The matrix $\mbb M(\phi) = M_1(\phi) M_2(\phi)$ is a product of the fermion matrices for each spin, which are given by
\beq \label{Mmatrix}
M_a(\phi) = \bbid + B_a(\phi_N) \cdots B_a(\phi_1),
\eeq
where $\phi_t =\{\phi_{t,x}\}$ denotes the field on time-slice $t$,
and the spatial matrices $B_a(\phi)$ are given by
\beq
B_a(\phi_t) = \me^{- H_2} \me^{- \tilde H_4(\phi_t)}
\eeq
with
\begin{align}
(H_2)_{x,y} & = \kappa \eps \delta_{\av{x,y}}  +  \varepsilon_a \mu \eps  \delta_{x,y}, \nn \\ 
\tilde H_4(\phi_t)_{x,y} & = -i \varepsilon_a \sin\phi_{t,x} \delta_{x,y}.
\end{align}
Here, $\delta_{\av{x,y}}$ is the nearest neighbor hopping matrix, and $\veps_1=+1, \; \veps_2=-1$. The parameter $\beta$ is related to $\epsilon U$ by
\beq
\me^{-\epsilon U/2} = 
\frac{I_0(\sqrt{\beta^2-1})}{I_0(\beta)}.
\eeq This path integral representation, derived in the appendix, is accurate up to terms of order $\epsilon^2$. Another representation, accurate only up to order $\epsilon$ is also discussed there.

\section{Infinite variance problem}

Consider the expectation value in \eq{eq:expectation}. For a fermionic observable, $\mathcal{O}(\phi) \det\mbb M(\phi)$ arises from the integration of a polynomial in Grassmann variables, resulting in a polynomial in the matrix elements of $\mbb M(\phi)$, and is therefore non-singular for a finite system. This remains true even for so-called ``exceptional configurations'' with $\det \mbb M(\phi) = 0$. It is also typically the case that $\mathcal{O}(\phi) \det\mbb M(\phi)$ does not vanish on these configurations.
For nearly-exceptional configurations with small but nonvanishing determinant, $\mbb M^{-1}$ exists, so $\mathcal{O}(\phi) \sim 1/\det\mbb M(\phi)$ can be large. While the contribution of such configurations to the average value of the observable is finite, their contribution to the variance $\sigma^2_{\mcal O} = \avg{\mcal O^2} -  \avg{\mcal O}^2$ is unbounded. In fact, the integrand in the numerator of
\bea
\langle \mathcal{O}^2(\phi)\rangle = 
\frac{
\int [\dd\phi] \mathcal{O}^2(\phi) \; \me^{-S_0(\phi)} \det\mbb M(\phi)
}
{\int [\dd\phi] \; \me^{-S_0(\phi)} \det\mbb M(\phi) 
}
\eea
is proportional to the divergent quantity $\mathcal{O}^2(\phi)  \det \mbb M(\phi) \sim 1/\det \mbb M(\phi)$. And because the polynomial $\mcal O \det \mbb M$ typically does not vanish at exceptional configurations, their presence will cause the variance to diverge.

\begin{figure*}[t!]
\includegraphics[width=0.50\textwidth]{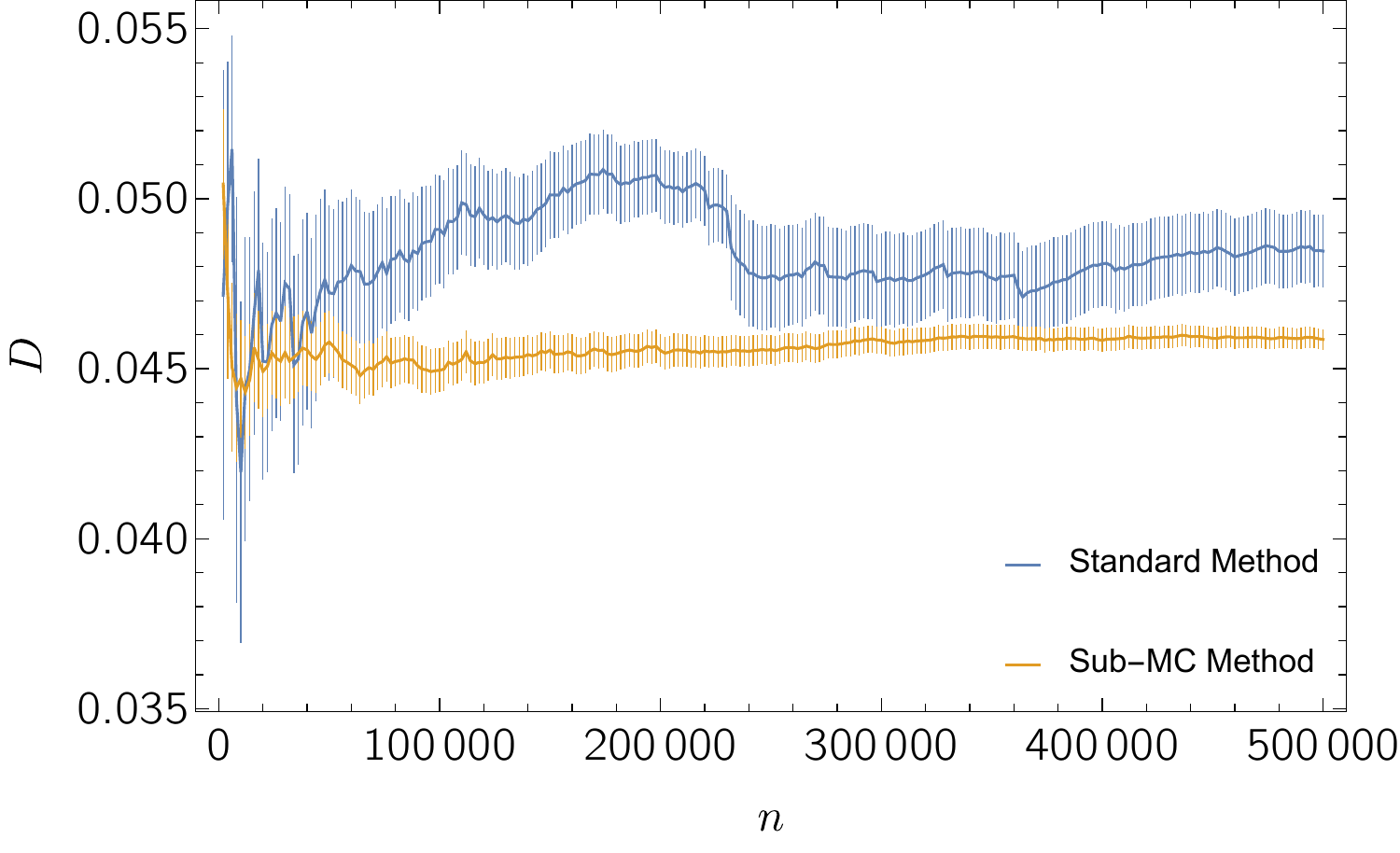}
\includegraphics[width=0.4825\textwidth]{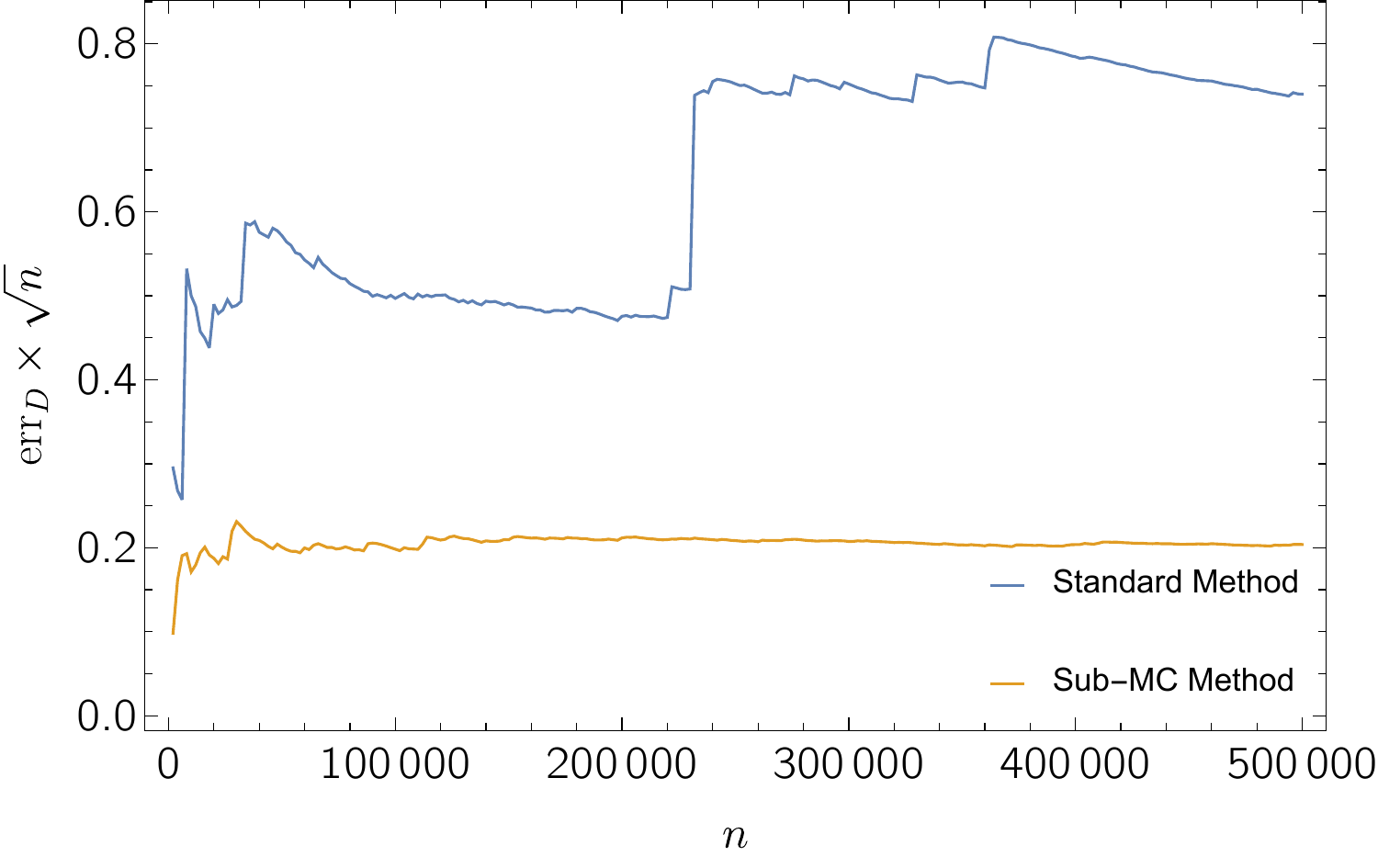}
\caption{\small{The double occupancy and its error estimate on a $4\times4$ lattice, with $U/\kappa=8$, $\mu=0$, $T/\kappa=0.5$, and $\eps=2.0$. The left plot is the Monte-Carlo history of the cumulative mean and error for the double occupancy as a function of number of updates, and the right plot is the scaled error, $\text{err}_D \times \sqrt{n}$, which should be constant if the error estimate is reliable. Sudden jumps in the error are a symptom of the infinite variance problem. The results here are produced using the improved action.
 \label{fig:infvariance}}}
\end{figure*}

\begin{figure}[b!]
\includegraphics[width=0.49\textwidth]{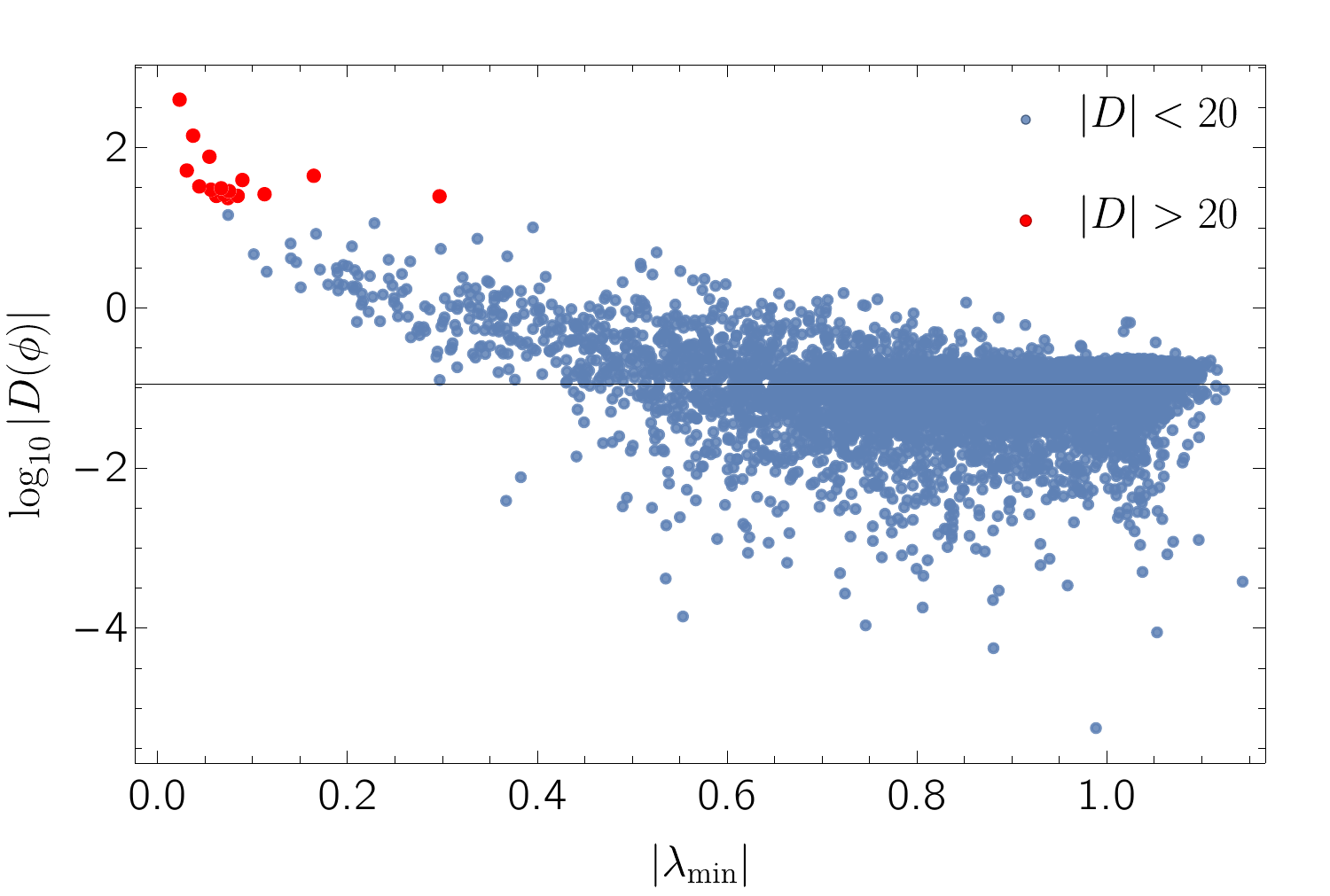}
\caption{\small{Double occupancy versus the smallest absolute eigenvalue of the fermion matrix $M(\phi)$. We highlight with red the ``exceptional'' configurations for which the double occupancy exceeds 20. The horizontal line marks the average value of $D$. \label{fig:exceptionals}}}
\end{figure}

In the Monte Carlo simulation, one computes a stochastic estimator $\overline{\mcal O}$ of $\avg{\mcal O}$, together with an error estimate given by the standard deviation of the mean
\beq
{\rm err}_{\overline{\mcal O}} = \sqrt{\frac{\sigma_{\mcal O}^2}n} = \sqrt{\frac{1}{n-1} \big( \overline{\mcal O^2} - \overline{\mcal O}^2 \big)},
\eeq
where $n$ is the number of statistically independent $\phi$ configurations generated in the Markov chain. When the variance $\sigma^2_{\mcal O}$ is finite, the error estimate ${\rm err}_{\overline{\mcal O}}$ decreases as $1/\sqrt{n}$. If the probability distribution is importance-sampled according to the distribution $p(\phi)\sim \exp[-S_0(\phi)] \det\mbb M(\phi)$, the exceptional configurations with vanishing $\det\mbb M(\phi)$ are never accepted, yet nearby configurations can be accepted, and these contribute arbitrarily large values for $\mathcal{O}^2(\phi) \det\mbb M(\phi)$, leading to the non-convergence of the variance estimator as $n$ is increased \cite{Yunus:2022pto, Shi:2016}. The effect of the infinite variance in a Monte Carlo calculation is exemplified by the upper curve in \fig{fig:infvariance}. Sudden ``glitches" caused by the sampling of a nearly-exceptional configuration change abruptly the average value of the observable (the double occupancy $D$ in this case) and lead to the non-convergence of the estimated error. In \fig{fig:exceptionals} we show that the large values of $D$ are correlated with small eigenvalues of the fermion matrix $M_a(\phi)$.

It would be advantageous to sample auxiliary fields according to a different statistical distribution where i) the variance with respect to the new distribution is not infinite, and ii) where the probability of the nearly-exceptional configurations is not so small and they are sampled more frequently while, at the same time, reweighting them down so their contribution is correctly taken into account. One such method is discussed in the next section.

\section{Methods}
\label{sec:methods}


\subsection{Extra time-slice reweighting}

To eliminate the infinite variance problem in our simulation, we follow the method proposed by Shi, et al. \cite{Shi:2016}. The basic idea is to sample the auxiliary field according to a modified distribution and reweigh the contribution of each configuration appropriately. We will choose the probability distribution of an identical system with {\it one extra} time-slice to define the distribution we will sample from. 
The rationale is that for a given auxiliary field this probability distribution, when neglecting the fields on the extra time-slice (marginalized distribution), is similar to the original probability density, but it has no zeros.

Formally, we multiply and divide the integrand of the partition function by a function $F(\phi)$ defined by 
\beq \label{F-def}
F(\phi) \equiv \int \dd\phi^* \; \me^{-S_0(\phi^*)} \det \mbb M_{N+1}(\phi,\phi^*),
\eeq
so that
\bea
Z &=&
\int [\dd\phi]_N \me^{-S_0(\phi)} \det\mbb M_N(\phi)
\frac{F(\phi)}{F(\phi)} \\
&=&
\int [\dd\phi]_N \dd\phi^* \;  R(\phi) \; \me^{-S_0(\phi,\phi^*)}
\det \mbb M_{N+1}(\phi,\phi^*), \nn
\eea
where we have defined a reweight factor $R(\phi) \equiv \det \mbb M_N(\phi) / F(\phi)$. Above, $S_0(\phi^*) = - \beta\sum_x \cos\phi^*_{x}$ is the bosonic action of the extra time-slice, and $S_0(\phi,\phi^*) = S_0(\phi)+S_0(\phi^*)$ is the total auxiliary field action over $N+1$ time-slices. For insertions of an observable $\mcal O(\phi)$, we similarly find
\bea
&&\int [\dd\phi] \me^{-S_0(\phi)} \mathcal{O}(\phi) \det\mbb M_N(\phi) \\
&=&
\int [\dd\phi]\dd\phi^* \mcal O(\phi) R(\phi) \; \me^{-S_0(\phi,\phi^*)}
\det\mbb M_{N+1}(\phi,\phi^*). \nn
\eea 
Observables may then be written as
\beq
\langle \mathcal{O}\rangle_N
=
\frac{\langle \mathcal{O}(\phi)  R(\phi) \rangle_{N+1}}{\langle R(\phi) \rangle_{N+1}},
\label{eq:rew}
\eeq where the symbol $\langle \cdots \rangle_{N+1}$ stands for the average with respect to the probability distribution of a system with an extra time-slice:
\beq
p_{N+1}(\phi,\phi^*) = 
\frac
{
\me^{-S_0(\phi,\phi^*)} \det\mbb M_{N+1}(\phi,\phi^*)
}
{\int[\dd\phi]\dd\phi^*
\me^{-S_0(\phi,\phi^*)} \det\mbb M_{N+1}(\phi,\phi^*)
}
\eeq
where $\det \mbb M_{N+1}(\phi,\phi^*)$ is the product
\BE
\prod_{a=1,2} \det[\bbid + B_a(\phi^*) B_a(\phi_N) \cdots B_a(\phi_1)].
\EE
At half-filling, these two determinants are complex conjugates of each other, and therefore $\det \mbb M_{N+1}$ is greater than or equal to zero. The function $F(\phi)$ is then strictly positive, since it is an integral of a non-negative quantity that cannot be uniformly zero, and because $\det \mbb M$ is bounded above on a finite lattice, $F(\phi)$ is also non-singular. This makes the reweight factor $R(\phi)$ strictly positive.

The infinite-variance problem is solved by this reweighting prescription because, if $\phi_0$ is an exceptional configuration  where $\det\mbb M_N(\phi)$ vanishes, then any positive power of $\mcal O(\phi_0) R(\phi_0)$ will be non-singular, since the determinant $\det \mbb M_N(\phi)$ cancels in their product. In particular, the integrand of the variance of $\mcal O R$ is $(\mcal O R)^2 p_{N+1}$, which is non-singular.

The computation of $F(\phi)$ can proceed in different ways. As described in Appendix \ref{sec:improved}, $F$ can be approximately written as a Grassmann integral of the observable $\exp(-\eps H)$ in the presence of a background field $\phi$. Perhaps the simplest way to compute it is then to expand in~$\eps$ as $\exp(-\epsilon H) = \bbid - \epsilon H + \ldots$ and compute the fermion contraction of $H$, and higher order terms, and truncate at some order. As we describe in Section~\ref{sec:results}, we find empirically that the $O(\epsilon^2)$ error introduced by this truncation is numerically very large. Instead, we will perform a non-perturbative Monte Carlo estimate of the reweighting factor by noticing that $F(\phi)$ can be computed as an expectation value
\beq
F(\phi) = {\cal Z}\av{\det \mbb M_{N+1}(\phi,\phi^*)}_g
\eeq
with respect to the probability distribution
$g(\phi^*)= \exp[-S_0(\phi^*)]/{\cal Z}$. Note that the normalization constant ${\cal Z}=\int \dd\phi^* \exp[-S_0(\phi^*)]$ can be evaluated analytically, but it can be neglected since it drops out in the ratio in Eq.~\ref{eq:rew}. This ``sub-Monte Carlo'' is relatively inexpensive, first because the distribution function $g(\phi^*)$ factorizes site-by-site and then standard methods can be used to sample efficiently the one-dimensional probability distribution.
Second, the matrices appearing in $\mbb M_{N+1}(\phi,\phi_*)$ have $\phi$ as a fixed background, so the matrix product $B_a(\phi_N)\cdots B_a(\phi_1)$ need only be computed once in the evaluation of $F(\phi)$.

\begin{figure*}[t!]
\includegraphics[width=0.49\textwidth]{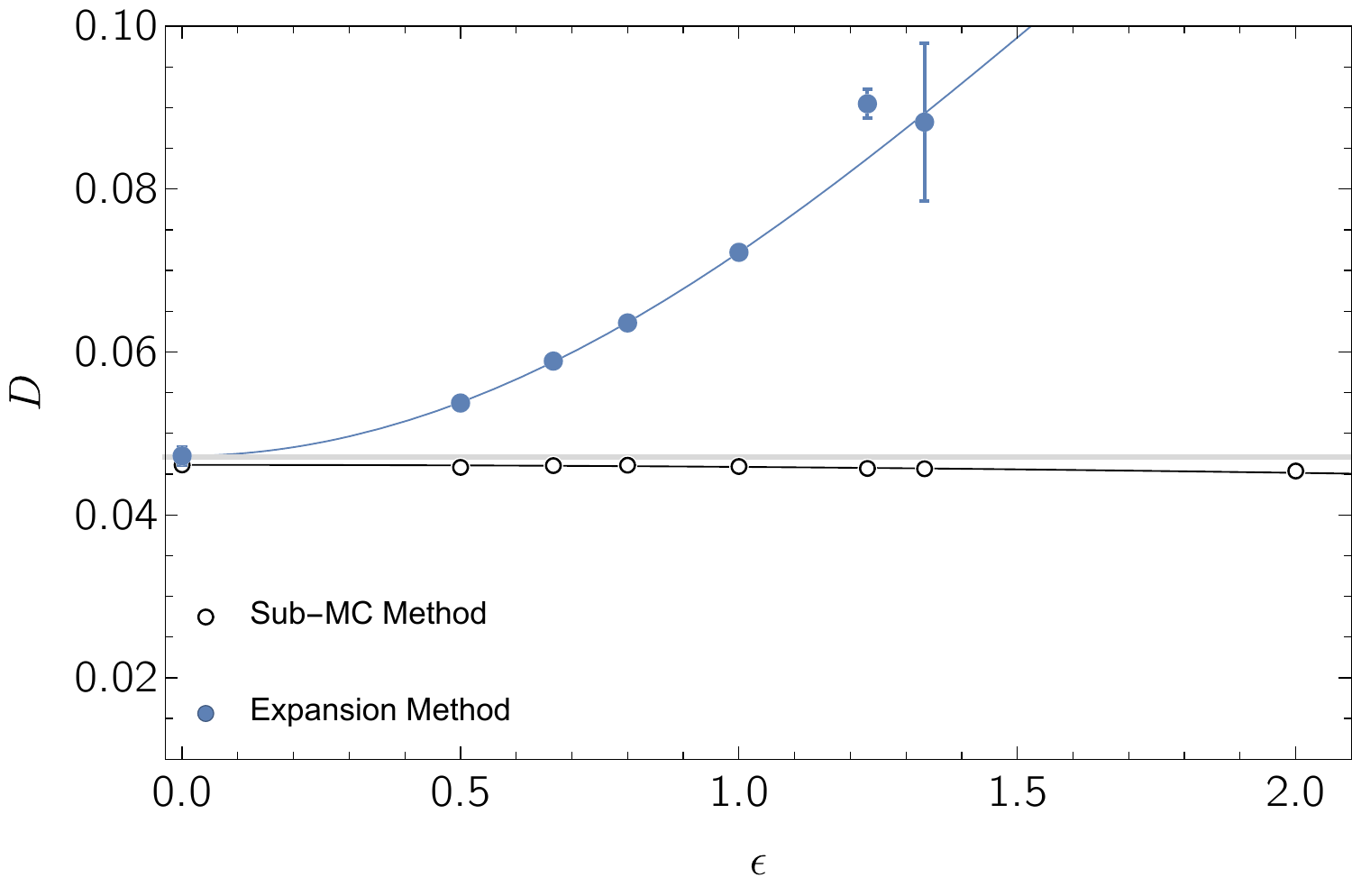}
\includegraphics[width=0.49\textwidth]{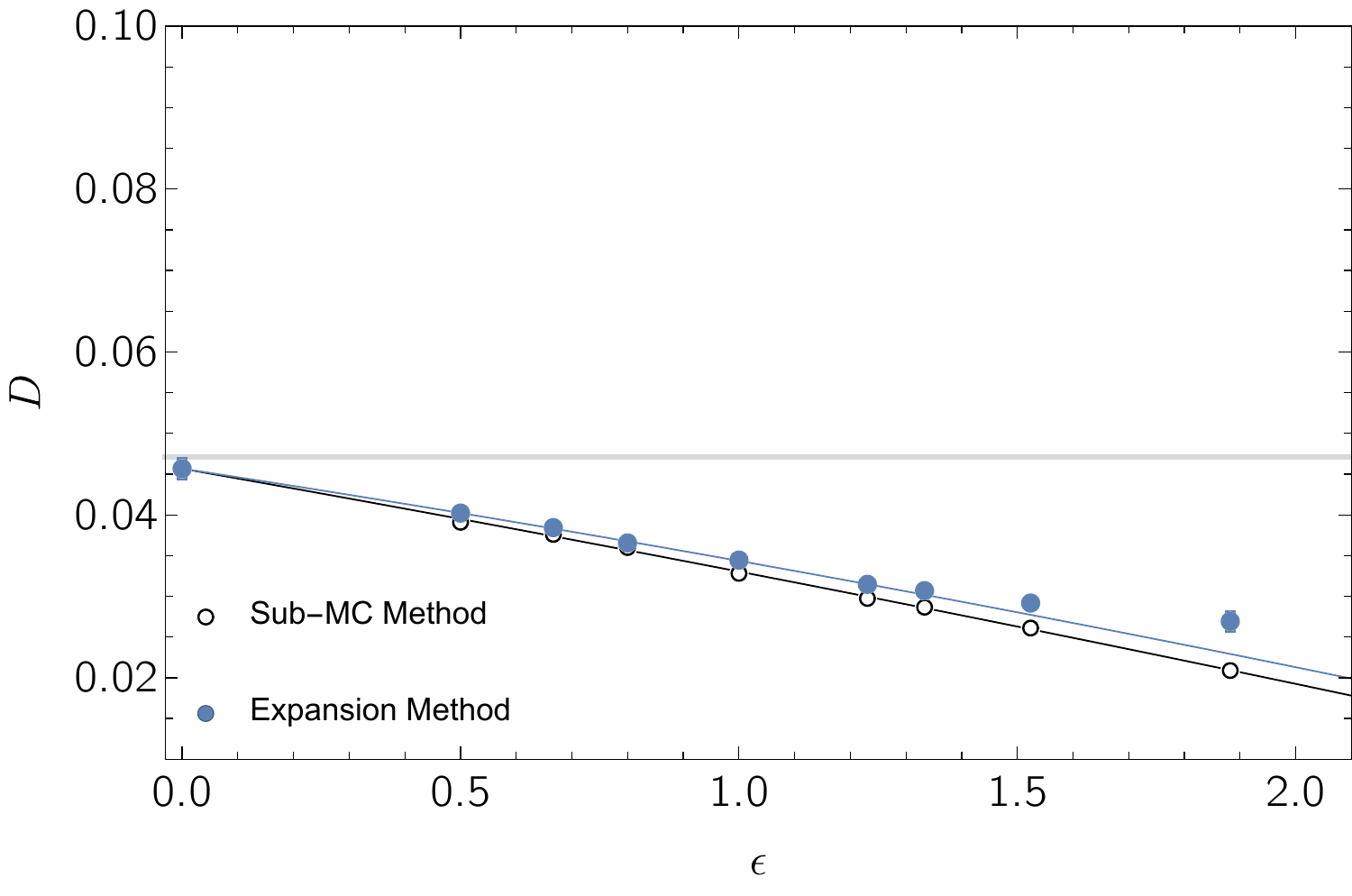}
\caption{\small{The double occupancy on a $4\times4$ lattice, with $U/\kappa=8$, $\mu=0$, and $T/\kappa=0.5$. The left and right plots are with the improved and conventional actions, respectively. The double occupancy is fitted as a function of $\eps.$ For the expansion method, the reweighting factor was calculated to order $\eps$ and only the three leftmost points are used in the fit. The gray band is the benchmark value of the double occupancy in infinite volume \cite{Benchmark:2015}. \label{fig:do}}}
\end{figure*}

\subsection{Unbiased estimator for $1/F$}

A subtlety arises in the Monte Carlo evaluation of $1/F(\phi)$. The Monte-Carlo estimator for $F(\phi)$ is computed by taking the mean over $n^*$ auxiliary variables $\phi^*_i$:
\beq
\overline{F(\phi)}\equiv{\cal Z} \frac1{n^*} \sum_i\det \mbb M_{N+1}(\phi,\phi^*_i)\, .
\eeq
$\overline{F(\phi)}$ is an unbiased estimator of $F(\phi)$, but $1/\overline{F(\phi)}$ is not an unbiased estimator of $1/F(\phi)$.

Lets denote ${\mathscr A}_i={\mcal Z}\det \mbb M_{N+1}(\phi,\phi^*_i)$, to simplify notation. We can compute the bias of the estimator $1/\overline{\mathscr A}$ from an expansion in $\mathscr A - \avg{\mathscr A}$,
\begin{equation}
    \begin{aligned}
       \Big\langle \frac{1}{\overline{\mathscr A}} \Big\rangle = \frac{1}{\avg {\mathscr A}}  -
       \Big\langle \frac{\overline {\mathscr A}-\avg {\mathscr A} }{\avg {\mathscr A}^2}
       \Big\rangle
        +
        \Big\langle \frac{ (\overline{\mathscr A}-\avg {\mathscr A})^2}{\avg {\mathscr A}^3}
        \Big\rangle
        +\cdots\,.
    \end{aligned}
\end{equation}
The second term vanishes but the remaining terms do not, which 
gives us a perturbative expansion for the bias.
In a Monte Carlo simulation, such a bias is only relevant if it is larger than the statistical error. Since the bias falls like $1/n$ to leading order, while the error falls as $1/\sqrt{n}$, in general one can control for the bias 
by increasing the sample size. However, if generating a large ensemble is expensive, a more elegant solution 
is to construct an unbiased estimator instead.

We adopt the unbiased estimator proposed in \cite{Moka:2019} for the reciprocal $1/\av{\mathscr{A}}$, 
for positive definite $\mathscr A_i > 0$. The estimator is defined by
\begin{equation}
    \hat \xi_{\mathscr A} \equiv \frac{w}{q_n} \prod_{i=1}^n (1-w {\mathscr A}_i) \,.
\end{equation}
Here $n$ is a non-negative integer-valued random variable distributed according to a geometric distribution with success probability $p$. The probability for outcome $n$ is $q_n=(1-p)^n p$. To calculate this estimator we have to evaluate $n$ uncorrelated random samples $\mathscr A_i$.
The average number of random variables required to compute the estimator once is $\av{n}=1/p$. The estimator is unbiased for any value of $p$ and any value of $w<2/\av{\mathscr A}$. 
For a given $\phi$ we tune the values of $w$ and $p$, to minimize the variance of the estimator, using the following method~\cite{Moka:2019}: We take $k$ samples of $\mathscr A_i$ and compute the means $\overline{\mathscr A}$ and $\overline{\mathscr A^2}$. We choose $w$ and $p$ according to
\begin{equation}
    \begin{aligned}
        & w = \mrm{min}\Big\{ \frac{1}{k \overline{\mathscr A}}, \frac{\overline{\mathscr A}}{\overline{\mathscr A^2}}, \frac{1}{{\mathscr A}_\mrm{max}} \Big\}, \\
        & p = 1 - \Big[1- 2 w \overline{\mathscr A} + w^2 \overline{\mathscr A^2} \Big]^{\frac{1}{2}}.
    \end{aligned}
\label{eq:wp}
\end{equation}
where ${\mathscr A}_\mrm{max}$ is the largest ${\mathscr A}_i$. The total number of samples to get one estimate is then $k+\av{n}=k+1/p$. In our simulations, it turns out that for most cases $w=1/k\overline{\mathscr A}$ and then
$p\approx 1/k$, thus the total cost for one evaluations is $2k$, on average.


\section{Results}
\label{sec:results}

\begin{figure*}[t!]
\includegraphics[width=0.49\textwidth]{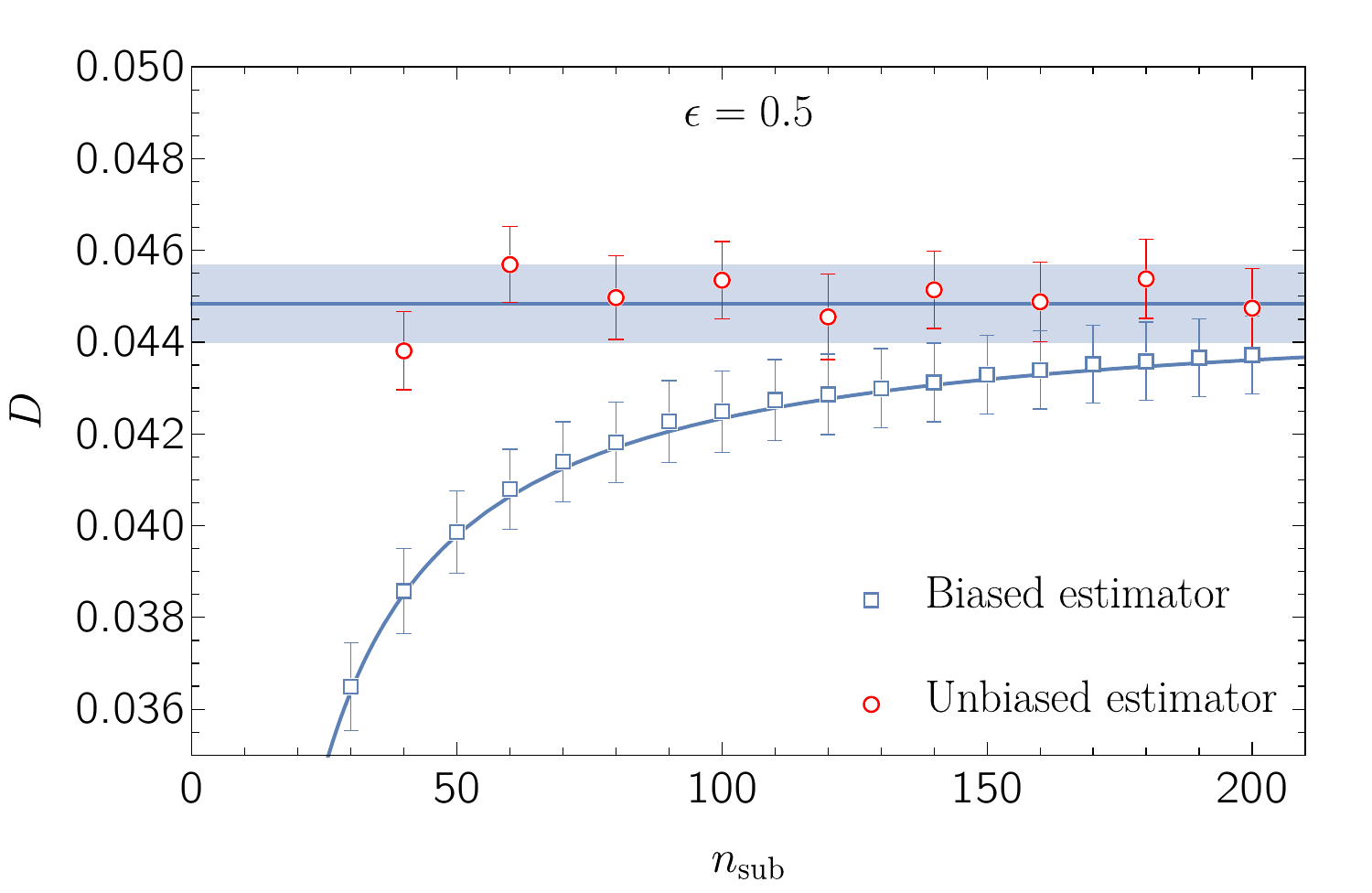}
\includegraphics[width=0.49\textwidth]{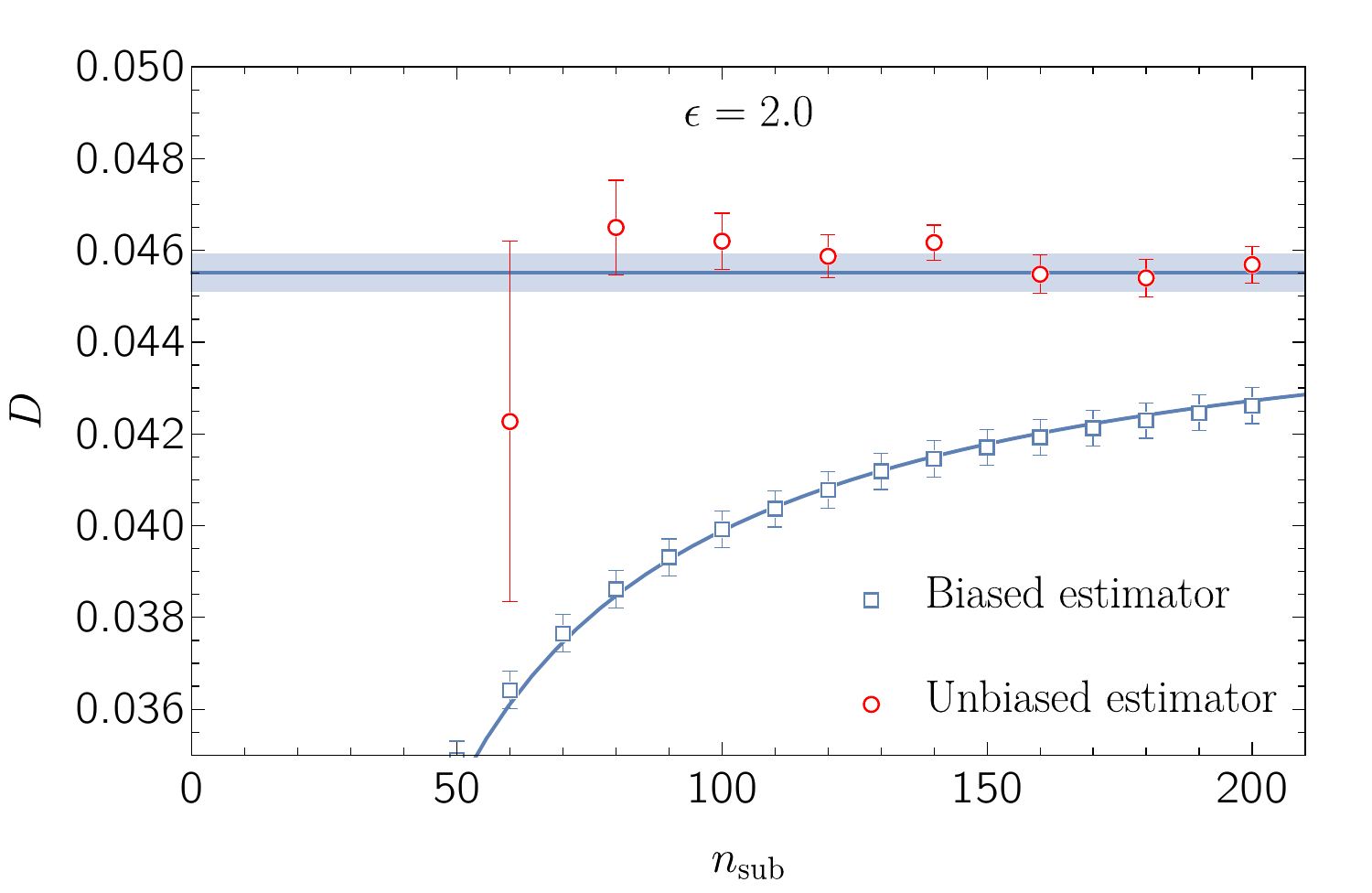}
\caption{\small{The double occupancy with biased and unbiased estimator on a $4\times4$ lattice, with $U/\kappa=8$, $\mu=0$, and $T/\kappa=0.5$. The left and right plots are at $\eps=0.5$ and $\eps=2.0$ respectively. The bias in the biased estimator vanishes as $1/n_\text{sub}$; the solid line represents that fit. The number of sub-MC samples for the unbiased estimator is random: here we use the average value $n_\text{sub}\approx 2k$, which includes the cost of the $k$ estimates required to compute $w$ and $p$ in eq.~(\ref{eq:wp}). The gray band is the extrapolation for the biased estimator at $n_\text{sub}\rightarrow \infty$ including its error bar. \label{fig:bias}}}
\end{figure*}

To test this method, we compute the double occupancy
\beqs
D(\phi) & = \frac{1}{V} \sum_x \av{n_\uparrow(x) n_\downarrow(x)}_F\\
& = \frac{1}{V} \sum_x [1-M_1^{-1}(\phi)_{x,x}]M_2^{-1}(\phi)_{x,x}
\eeqs
for Hubbard model at half-filling in a small volume. 
The infinite-variance problem is apparent in \fig{fig:infvariance}.
In the same figure, the corresponding calculation using the extra time-slice reweighting (with the unbiased estimator) is shown. It is evident that the glitches are gone or, at least, drastically reduced. Even more evident is the fact that the variance of the sampled mean approaches a constant instead of diverging as the number of samples is increased. This difference in behavior between the calculation with and without reweighting was consistently observed over a range of Hubbard model parameters and for different variations of the reweighting procedure (which are described next).

In the left pane of \fig{fig:do} we show the result of the $\epsilon\rightarrow 0$ extrapolation using the ``improved" action (accurate to $O(\epsilon^2)$) but computing the reweighting factor $R(\phi)$ in two ways: using sub-Monte Carlo estimation and from an expansion of $\exp(-\eps H)\approx 1-\eps H$ that introduces $\eps^2$ corrections. The sub-Monte Carlo method has a mild $\eps$-dependence; this may be expected from the fact that the improved action is correct to order $\eps^2$ and that no further approximation is made in computing $R$, but we also observe that the coefficient of $\eps^2$ is small. Meanwhile, the $\exp(-\eps H)$ expansion, while parametrically of the same order as the sub-Monte Carlo method, introduces a notable systematic error, shifting the $\eps^2$ coefficient substantially. In the right pane of \fig{fig:do}, the corresponding results computed from the ``conventional'' action described in Appendix~\ref{sec:conventional} are displayed. This action is accurate to order $\eps$, and this is reflected in the data; the systematic error in $\eps$ is large. We note that it is likely that higher order expansions of $\exp(-\eps H)$ will show a better continuum limit behavior than the $\eps^2$ truncation. However, the calculation of $R$ become quickly impractical due to the number of contractions needed to compute higher powers of $H$.

This indicates that if one wants to keep the advantages of using the improved action and, at same time, circumvent the infinite variance problem using the extra time-slice reweighting, the best strategy is to use the sub-Monte Carlo estimator, as described in the previous section. 

The usefulness of our method hinges on being able to estimate the reweighting factor with a moderate number of sub-Monte Carlo samplings. In \fig{fig:bias} we show the bias for the double occupancy as a function of the (average) number of sub-Monte Carlo samplings required.
We compare the unbiased estimator discussed above, with a biased one that uses $1/\overline F$ as an estimator for $1/\av{F}$. The results clearly show that the unbiased estimator is superior to the biased one: the error bars are comparable and the bias vanishes. We note that the stochastic errors on the final results are a combination of variance of the observable from configuration-to-configuration and the variance of the estimator. As we reduce the variance of the estimator, which comes at increasing cost, the errors are asymptotically converging to the value controlled by the configuration-to-configuration fluctuations. The  number of sub-Monte Carlo samplings required to reach the asymptotic regime turns out to be very modest and, since the reweighting is done only on the configurations used for measurements, the overhead of using our method is minimal.

\section{Discussion}
\label{sec:conclude}

Using the Hubbard model as a testbed, we  studied the infinite variance problem in a fermionic system and suggested a new way to improve the extra time-slice method proposed in Ref.~\cite{Shi:2016}. 
We established that computing the reweighting factor in powers of  the time step $\epsilon$, as advocated in Ref.~\cite{Shi:2016}, leads to large finite $\epsilon$ corrections, at least at leading order. We proposed instead to compute the reweighting factor using a sub-Monte Carlo calculation and pointed out that using an unbiased estimator allow us to use very large values of $\epsilon$.


To assess the effectiveness of our method, we calculated the double occupancy of Hubbard model. We used two different discretizations, one correct up to $O(\epsilon)$ and another, improved one, correct up to $O(\epsilon^2)$. In order to take full advantage of the convergence properties of the improved action it is essential that the reweighting is done non-perturbatively and using an unbiased estimator. 

We note that for efficient simulations, we need to use the improved action.
As shown in \fig{fig:do}, one can use larger values of $\epsilon$ with our improved action.
Larger values of $\epsilon$ translate in a smaller number of time-slices with a resulting lowering of computational costs.
In fact, the bottleneck in Monte Carlo simulation of fermionic theories is the calculation of the fermion determinant. This scales in general as $O(V^3N^3)$, for direct evaluations, and as $O(V^3N)$ when using the reduction formula as in eq.~(\ref{sausage}) where $V$ is the volume of spatial lattice and $N$ is the number of time-slices. Additionally, autocorrelation is generally increased at smaller values of $\eps$, which then favors simulations with large values of $\eps$.
On another hand, the improved action simulations exposed the infinite variance problem, which seems to be more prominent than when simulating the unimproved action.

The success of the extra time-slice method requires that $F(\phi)$ is strictly greater than zero, which holds for the Hubbard model at half-filling. Away from half-filling, a sign problem develops which then spoils this property, and a new definition of $F(\phi)$ will be required to avoid the variance problem. However, it may be possible to apply a modified extra time-slice method in that scenario; we will explore this in future work.

Lastly, we note the fact that $1/\overline{\mathscr A}$ is a biased estimator for $1/\avg{\mathscr A}$ also implies that in traditional applications of reweighting, for example in simulations with a sign problem, the estimate for $1/\avg{R}$ (see denominator in eq.~(\ref{eq:rew})) is biased. In fact, if the \textit{same} ensemble is used to compute the numerator and denominator, the overall estimator for any observable will have a bias different from that of $1/\overline{R}$ alone, and therefore care should be taken to define unbiased estimators for observables in reweighted theories, especially when the ensemble size is not large.




\begin{acknowledgments}

This work was supported in part by the U.S. De- partment of Energy, Office of Nuclear Physics under Award Number(s) DE-SC0021143, and DE-FG02- 93ER40762, and DE-FG02-95ER40907.

\end{acknowledgments}

\appendix
\section{Path integral representations}

\subsection{$O(\eps)$ discretization}\label{sec:conventional}

The first discretization of the partition function $Z= \mrm{Tr}[\exp(-\eps H)]^N$ is obtained by resolving the trace in terms of fermion coherent states~\cite{Montvay:1994}, and then expanding the exponential $\exp(-\eps H)$ in powers of~$\eps$. Then, upon inserting coherent state identities and using the property
\beq
\langle \eta | F(\hat{\psi}^{\dagger},\hat{\psi}) | \eta' \rangle = F(\bar\eta,\eta') \ \me^{\sum_{x,a}\bar\eta^a_x\eta'^a_{x}},
\eeq
for any polynomial function $F(\hat \psi^\dag, \hat\psi)$ in normal-ordered form, the partition function becomes a Grassmann integral over a product of factors
\beq \label{gman-sandwich}
\langle \eta_{t+1} | \me^{-\eps H} | \eta_{t} \rangle = \me^{\sum_{x,a}\bar\eta^a_{t+1,x} \eta^a_{t,x}} \Big(\me^{- \mcal H(\bar \eta_{t+1}, \eta_t)} + O(\eps^2) \Big),
\eeq
where $\mcal H = \mcal H_2 + \mcal H_4$, with $\mcal H_2$ quadratic in Grassmann variables and $\mcal H_4$ quartic:
\begin{align}
\mcal H_2 = \eps \kappa \sum_{\av{x,y},a} \bar \eta^a_{t+1,x} \eta^a_{t,y} + \sum_{x,a} Z_a \bar \eta^a_{t+1,x} \eta^a_{t,x}
\end{align}
with $Z_a = \smallfrac{\eps U}{2} + \veps_a \eps \mu, \; \veps_1=+1, \; \veps_2=-1$, and
\beq
\mcal H_4 = \frac{U \eps }{2} \sum_{x,a,b} \big(\bar\eta^a_{t+1,x} (\sigma_3)_{ab} \eta^b_{t,x}\big)^2.
\eeq
Here, $\sigma_3$ is the Pauli-$z$ matrix.

We perform a Hubbard-Stratonovich (HS) transformation to a compact auxiliary  field by noting that for any $X$  such that $X^2 \neq 0$ but $X^3 = 0$ we have,
\beq \label{HS-1}
\frac{1}{2\pi I_0(\beta)} \int_0^{2\pi} \!\! \dd\theta \; \me^{\beta \cos\theta} \me^{i X\sin\theta }=  \me^{-\frac{U\eps}{2} X^2},
\eeq
with $\beta$ determined from the condition $U \eps = I_1(\beta) / \beta I_0(\beta)$, and where $I_n(z)$ are modified Bessel functions. Thus, letting $X = \bar\eta^a_{t+1,x} (\sigma_3)^{ab} \eta^b_{t,x}$, the factors $\exp(- \mcal H_4)$ for each time-slice become integrals over $\phi_t = \{\phi_{t,x}\}$, and one can then perform the Grassmann integration to find
\beq
Z = \int [\dd\phi] \me^{-S_0(\phi)} \det\big(\mbb M_1(\phi) \mbb M_2(\phi)\big) + O(\eps)
\eeq
where $[\dd \phi] = \prod_{t,x} \dd \phi_{t,x}$, $S_0$ is the auxiliary field action, eq. (\ref{free-action}), and the fermion matrix is given by
\beq
\mbb M_a (\phi)_{(t,x),(t',y)} = \delta_{x,y} \delta_{t,t'} + A_a(\phi_t)_{x,y} b_t \delta_{t,t'+1}
\eeq
with
\beq
A_a(\phi_t)_{x,y} = (Z_a - 1 - \veps_a i \sin \phi_{t,x}) \delta_{x,y} - \eps \kappa \delta_{\av{x,y}}
\eeq
Here, $b_{t}=-1$ for $t=N-1$ and $1$ otherwise, due to antiperiodicity, and $\delta_{\av{x,y}} = \sum_{\mu} \delta_{x,y+\mu}$ is the hopping matrix. Using the following identity for an arbitrary set of matrices $\{A_t\}$ \cite{Blankenbecler},
\begin{equation} \label{sausage}
    \det 
    \begin{bmatrix}
        \bbid & 0 & \cdots & A_N \\
        A_1 & \bbid & \cdots & 0 \\
        \vdots & \vdots & \ddots & \vdots \\
        0 & \cdots & A_{N-1} & \bbid \\
    \end{bmatrix}
    = \det\Big(\bbid-(-1)^N A_N \cdots A_1 \Big)
\end{equation}
we can show that
\beq \label{conv-action}
Z = \int [\dd\phi] \me^{-S_0(\phi)}  \det M_1(\phi) \det M_2(\phi) + O(\eps)
\eeq
where 
\beq
M_a(\phi) = \bbid + (-1)^N A_a(\phi_N) \cdots A_a(\phi_1).
\eeq
The error in using this action is $O(\eps)$ because, in taking the product of eq. (\ref{gman-sandwich}) for all $t$, the subleading term is $O(N \eps^2) = T^{-1} O (\eps)$, $T$ being the temperature.

\subsection{$O(\eps^2)$ discretization}\label{sec:improved}
The second discretization is derived by a different Trotterization which separates quadratic and quartic terms in the Hamiltonian:
\begin{equation}
Z = \mrm{Tr}\big[(\me^{-\eps H_2} \me^{-\eps H_4})^N \big] + O(\eps^2)
\end{equation}
with
\begin{align}
        H_2 &= -\kappa \sum_{\av{x,y}} \sum_a \hat{\psi}^{\dagger}_{a,x}\hat{\psi}_{a,y} -\mu \sum_x (\hat{n}_{1,x} - \hat{n}_{2,x}) \nn \\
        H_4 &= \frac{U}{2} \sum_x (\hat{n}_{1,x} - \hat{n}_{2,x})^2 
\end{align}
Above, $\hat{n}_{1,x}$ is the number operator for up-electrons on site $x$ and $\hat{n}_{2,x}$ the number operator of down-holes.

We implement a HS transformation to a compact auxiliary field by noting that eq. (\ref{HS-1}) is also valid for any operator $\hat X$ with eigenvalues $\{0,\pm 1\}$,
with $\beta$ instead satisfying $\me^{-U \eps/2}=I_0(\sqrt{\beta^2-1})/I_0(\beta)$. If we now use the coherent state property \cite{Montvay:1994}, 
\begin{equation}
    \langle \eta | \me^{\sum_{x,y} \hat{\psi}^{\dagger}_{x}A_{xy}\hat{\psi}_{y}} | \eta' \rangle = \me^{\sum_{x,y} \bar \eta_{x} (\me^{A})_{xy} \eta'_{y}},
\end{equation}
and let $\hat X = \hat n_{1,x} - \hat n_{2,x}$, we can write the partition function as
\beq
    \begin{aligned}
        Z & = \int [\dd\phi] \; \me^{-S_0(\phi)} \det\big(\mbb M_1(\phi) \mbb M_2(\phi)\big) + O(\eps^2),
    \end{aligned}
\eeq
where the fermion matrix $\mbb M_a$ is now given by
\beq
\mbb M_a (\phi)_{(t,x),(t',y)} = \delta_{x,y} \delta_{t,t'} + B^a(\phi_t)_{x,y} b_t \delta_{t,t'+1}
\eeq
and
\beq
B^a(\phi_t)_{x,y} = \me^{- \tilde H_2^a} \me^{- \tilde H_4^a(\phi_t)}
\eeq
with the definitions
\begin{equation}
    \begin{aligned}
        & (H_2^a)_{x,y} = -\eps \kappa \delta_{<x,y>} +\varepsilon_a \eps \mu \\
        & \tilde H_4^a(\phi_t)_{x,y} = - i \varepsilon_a \delta_{x,y} \sin \phi_{t,x}.
    \end{aligned}
\end{equation}

Using eq. (\ref{sausage}) again, one arrives at
\begin{align} \label{imp-Z}
Z = \int [\dd\phi] \; \me^{-S_0(\phi)}  \det M_1(\phi) \det M_2(\phi) + O(\eps^2)
\end{align}
where
\beq \label{M-def}
M_a(\phi) = \bbid + B^a(\phi_N) \cdots B^a(\phi_1).
\eeq
We emphasize that $\beta$ here is a different function of $U\eps$ than the one appearing in eq. (\ref{conv-action}). Furthermore, the discretization error in this expression for the partition function is $O(\eps^2)$, whereas for the conventional action in eq. (\ref{conv-action}), the error was $O(\eps)$. In this sense, the action in eq. (\ref{imp-Z}) is ``improved.''

The ``expansion method'' in \fig{fig:do} refers to an alternative way of evaluating $F(\phi)$, eq. (\ref{F-def}), in terms of fermion contractions with respect to an action over $N$ time-slices. First begin with the identity
\BE
\det \mbb M_{N+1}(\phi,\phi^*) = \mrm{Tr} \big[ B(\phi^*) B(\phi_N) \cdots B(\phi_1) \big],
\EE
where $B(\phi) = B_1(\phi) B_2(\phi)$. Integrating over $\phi^*$ with a density $\me^{-S_0(\phi^*)}$, we obtain
\begin{align}
F(\phi) & = \mrm{Tr} \big[ \me^{- \tilde H_2} \me^{-\eps H_4} B(\phi_N) \cdots B(\phi_1) \big] \\
& = \mrm{Tr} \big[ (1 - \eps H ) B(\phi_N) \cdots B(\phi_1) \big] + O(\eps^2). \nn
\end{align}
The quantity in the trace can be evaluated using the properties of Grassmann coherent states as
\BE
G_1(\phi) = \big( 1 - \eps h_1(\phi) \big) \det \mbb M_N(\phi)
\EE
where
\begin{align}
h_1(\phi) = & \; \kappa \sum_{\av{x,y}}\big( (M^{-1}_1)_{y,x} + (M^{-1}_2)_{y,x} \big) \nn\\ 
& + \frac{U}{2} \sum_x \Big( (M^{-1}_1)_{x,x} + (M^{-1}_2)_{x,x} \nn\\
& \quad\quad\quad\quad - 2 (M^{-1}_1)_{x,x} (M^{-1}_2)_{x,x} \Big) \nn \\
& + \mu \sum_x \big(  (M^{-1}_1)_{x,x} - (M^{-1}_2)_{x,x} \big) 
\end{align}
and $M^{-1}_a(\phi)$ is the matrix inverse of eq. (\ref{M-def}). Thus $F(\phi) = G_1(\phi) + O(\eps^2)$. Higher orders may be computed, but the number of fermion contractions grows exponentially.

\bibliography{hubbard_refs.bib}

\end{document}